\documentclass{aastex}
\usepackage{spr-astr-addons}
\usepackage{url}

\begin{document}

\title{The fallacy of Oppenheimer Snyder Collapse:  no general relativistic Collapse at all, no black hole, no physical singularity} 
\shorttitle{No uniform Density Star}
\shortauthors{A. Mitra}

\def\b{\begin{equation}}
\def\e{\end{equation}}
\def\l{\left}
\def\r{\right}

\author{Abhas Mitra\altaffilmark{1}}

\altaffiltext{1}{Theoretical Astrophysics Section, Bhabha Atomic Research Centre,
    Mumbai -400085, India: Email: amitra@barc.gov.in}

\begin{abstract}
By applying Birkhoff's theorem to the problem of the general relativistic collapse of a uniform density dust, we directly show that the density of the dust $\rho=0$ even when its proper number density $n$ would be assumed to be finite! The physical reason behind this exact result can be traced back to the observation of Arnowitt et al. that the gravitational mass of a neutral point particle is zero: $m=0$ (PRL, 4, 375, 1960). And since, a dust is a mere collection of {\em neutral point particles, unlike a continuous hydrodynamic fluid}, its density $\rho = m n=0$. It is nonetheless found that for $k=-1$, a homogeneous  dust can collapse and expand special relativistically in the fashion of  a  Milne universe. Thus, in reality, general relativistic homogeneous dust collapse does not lead to the formation of any black hole in conformity of many previous studies (Logunov,  Mestverishvili,   Kiselev, Phys.Part.Nucl. 37, 317, 2006; Kisevev, Logunov \& Mestvirishvili, Theor. Math. Phys., 164, 972, 2010;
Mitra, J. Math. Phys.  50, 042502, 2009; Suggett, J. Phys. A, 12, 375 1979).
      Interestingly, this result is in agreement with the intuition of Oppenheimer \& Snyder (Phys. Rev. 56,  p.456, 1939) too: 

``Physically such a singularity would mean that the expressions used for the energy-momentum tensor does not take into account  some  essential physical fact which would really  smooth the singularity out. Further, a star in its early stages of development would not possess a singular density or pressure, it is impossible for a singularity to develop in a finite time.''

\end{abstract}
\

\keywords{Gravitational Collapse; Oppenheimer \& Snyder Collapse; Black Hole}
\maketitle


\section{Introduction}
 In view of the virtual impossibility of having {\em exact}  analytical solutions of general relativistic gravitational collapse problem, Oppenheimer and Snyder (1939) assumed the collapsing star fluid to be not only homogeneous but also a ``dust'', i.e., having no pressure at all $p=0$. They justified this extreme assumption on the plea that at the exhaustation of nuclear fuel and lack of energy generation, pressure of the star would (almost) drop to zero. Then they treated the problem by considering both a non-comoving as well as comoving frame. And while considering the latter, they eventually considered  what in modern terminology is called a ``marginally bound dust', i.e., one having a Newtonian energy per unit mass, $E=0$ ($k=0$). 

 If despite assuming a strict $p=0$ equation of state (EOS) of the collapsing fluid, its density would be assumed to be finite,
then OS solution would suggest the formation of a black hole (BH) in a finite comoving proper time $t_c = (6\pi \rho_0)^{-1/2}$, where $\rho_0$ is the initial density. OS also claimed to have shown that an event horizon would form at an {\em unspecified} moment so that this collapse would indeed appear to form a black hole. Subsequently, innumerable authors, innumerable articles have reexamined this problem by making the same assumption that despite a $p=0$ EOS of the collapsing fluid, its density can not only be finite, but even infinite too \citep{b28}. Such studies have established the generic nature of the OS work; that even if one would consider bound ($E<0; k=+1$) or unbound ($E >0; k=-1$) homogeneous dust, one would still find a black hole. 

We  however emphasize the already known fact that, in the comoving frame, the results of the OS collapse matches {\em exactly} with their Newtonian counterparts \citep{b6}.   For an exact Newtonian result, one needs to have a value of Kretschmann scalar $K \to 0$. And this is possible only when all components of the energy momentum tensor $\to 0$. For a dust, one already has $p=0$;
and one gets a hunch, that the $p=0$ equation of state (EOS) may indeed require $\rho=0$ too so that one can have $K \to 0$!  

To resolve this issue, we shall invoke Birkoff's theorem by which the exterior spacetime of any adiabatically evolving spherical fluid can be represented by the so-called vacuum Schwarzschild solution. Then it would be found that the homogeneous dust necessarily has $\rho=0$ irrespective of whether it is contracting or expanding. We would explain this exact result by banking on the ADM result that the {\em gravitational mass of a neutral point particle is zero}: i.e., $m=0$  for a dust particle\citep{b1,b2}.

 We would also critically analyze the mental picture about the inevitability of the dust collapse. Since it is found that $\rho=0$, the spacetime associated with the OS problem would be the flat Minkowski one. And it would be seen that an unbound $k =-1$ dust may undergo a notional/mathematical  collapse to the center of symmetry in a finite proper time and form a density caustic having $n=\infty$.  Further, in a absence of any trapped surface or spacetime singularity, this collapsing dust should overshoot the caustic to reemerge as an expanding dust.  In this expanding form, the OS dust ball would be seen to form the special relativistic universe conceived by Milne\citep{b8, b13}.

 Finally, we shall try to appreciate the entire picture from a physical perspective.

\section{Collapse in a comoving frame}

  The general spherically symmetric metric in comoving coordinates is given by ($G=c=1$)
 
 \begin{equation}
 ds^2 = e^{{\nu}(r,t)} dt^2 - e^{{ \lambda}(r,t)} dr^2 - R^2 d\Omega^2 
 \end{equation}

 where $d \Omega^2 = (d\theta^2 + \sin^2\theta d\phi^2)$ and $R(r,t)$ is the curvature coordinate. In the comoving frame,
  the components of the perflect fluid stress-energy tensor are
 \begin{equation}
 T_1^1(com) =T_2^2(com) = T_3^3(com) = p;  T_0^0(com) = -\rho
 \end{equation}

 One important parameters here is the  Misner-Sharp mass\citep{b14}
 \begin{equation}
 M(r,t)  = \int_0^r 4 \pi \rho R^2 R' dr,
 \end{equation}

For a uniform density case with $\rho=\rho(t)$, the above expression becomes
\b
M(r,t) = {4\pi \rho(t)\over 3} R^3 - M(0,t)
\e
And in order that the spacetime is regular at $r=0$, one must have $M(0,t) =0$ so that, for a uniform density case, one has
 \b
M(r,t) = {4\pi \rho(t)\over 3} R^3 
\e

Note, this may also be written as
\b
M(r,t) = \int_0^R 4 \pi \rho(t) R^2 dR
\e

 Thus at the boundary of the fluid, one will have 

\b
M_b(t) = \int_0^{R_b} 4 \pi \rho(t) R^2 dR
\e

\section{Collapse in Noncomoving Coordinates}
One may study the problem of gravitational collapse in non-comoving coordinates too. In fact OS started their investigation by using the following metric:
\b
ds^2 = e^{\eta} dT^2 - e^{\Lambda} dR^2 - R^2 d\Omega^2
\e
In the comoving frame, the clocks have $r=fixed$ and the fluid is at rest. But in the non-comoving frame considered above, the clocks have $R=fixed$ and the fluid is in  motion.

  The 3-speed $v$ of the fluid as measured by this non-comoving frame  is given by\citep{b11} 
\b
v^2 = {e^{\Lambda} dR^2\over e^{\eta} dT^2} 
\e

In this non-comoving coordinate system, one has \citep{b11}

\b
8 \pi T_0^0 = e^{-\Lambda} \l({1\over R^2} - {\Lambda'\over R}\r) - {1\over R^2}
\e
where a prime denotes partial derivative by appropriate radial coordinate and
\b
T_0^0 = -{\rho + p v^2\over 1-v^2}
\e
And for a dust with $p=0$, one has
\b
T_0^0 = -{\rho\over 1-v^2} 
\e
By integrating Eq.(10),  one obtains
\b
e^{-\Lambda} = 1 - {2{\cal M} \over R}
\e
where
\b
{\cal M} = -\int_0^R 4 \pi T_0^0 R^2 dR =\int 4 \pi {\rho \over 1-v^2} R^2 dR
\e

\section{Application of Birkhoff's Theorem}
As we know, by Birkoff's theorem, the exterior vacuum spacetime of not only a static sphere but also that of an adiabatically evolving sphere, is given by the vacuum Schwarzschild solution:
 \b
ds^2 = e^{\eta_e} dT^2 - e^{\Lambda_e} dR^2 - R^2 d\Omega^2
\e
where
\b
e^{\eta_e} = e^{-\Lambda_e} = 1 - {2M_b\over R}; \qquad R \ge R_b
\e
On the other hand, from the {\em interior solution} Eq.(13), we find that
\b
e^{-\Lambda_b} = 1 - {2 {\cal M}_b\over R_b}
\e
Since the interior and exterior metric must match at the boundary $R=R_b$, from the two foregoing equations we find
\b
e^{-\Lambda_b} = 1 - {2 {\cal M}_b\over R_b}= 1 -{2M_b\over R_b}
\e
   Therefore, for any adiabatically evolving sphere, we obtain
\b
M_b= {\cal M}_b 
\e
This means for a uniform density collapsing/expanding dust sphere
\b
\int_0^{R_b} 4 \pi \rho(t) R^2 dR = \int_0^{R_b} 4 \pi R^2 {\rho(t) \over 1 -v^2} dR
\e
By transposing, we find

\b
{\cal M}_b - M_b =0 = \int_0^{R_b} 4 \pi R^2 \l[ {\rho \over 1 -v^2} - \rho\r]dR
\e
or,
\b
\int_0^{R_b} 4 \pi R^2 {\rho(t) v^2\over 1 -v^2} dR =0
\e

Since $1-v^2 >0$ in the foregoing Eq., it can be satisfied only if
\b
\rho(t) v^2 =0
\e
And for an assumed dynamic problem, this means that
\b
\rho(t) =0
\e

\section{Physical Explanations}

A ``dust'' is a collection of countable individual {\em neutral point particles}. This is quite unlike the concept of a physical fluid with {\em smeared out} matter distribution. While in the latter case, one can speak of only a smeared out fluid element, for a dust, one can talk in terms of individual {\em neutral point particles}. Thus, the proper mass density of a dust is $\rho =\rho_b= n m$, where $\rho_b$ is the proper rest mass density. In contrast for a hydrodynamic fluid, $\rho \neq \rho_b$. Let us again try to appreciate why the dust particles must indeed be geometrical points:

If a dust ``particle'' would be conceived of as a finite mass sphere with a finite radius, a given sphere would be subject to tidal pull by the interior dust. This tidal pull would immediately elongate and distort the given ``sphere''. The gravitation interaction due to such a distorted sphere would disturb the preexisting harmonious dust motion. In other words, the dynamic gravitational field of distorted spheres would {\em induce random motions} amongst themselves. If so, the strict $p=0$ condition would be violated. Thus in order to ensure that an individual ``dust particle'' is not subject to  any tidal distotion, it must be a geometric point. On the other hand for a smeared out hydrodynamic fluid, a given shell comprises a continuous fluid rather than {\em countable individual} particles. And when the shell itself would be subjected to tidal pull by the interior fluid, spherical symmetry would not be affected.

  Having expounded on this {\em subtle point}, now we may  recall
 that long back  ADM\citep{b1}  found that the {\em gravitational mass of a neutral point particle} of radius $\epsilon \to 0$ is {\bf zero}. Let us reproduce the Eq.(4) of their paper:
\b
m= \lim_{\epsilon \to 0} 2m_0 \l[1 + (1+m_0/8\pi \epsilon)^{1/2}\r]^{-1}
\e
where $m_0$ is the bare mass and $m$ is the total mass ($E=mc^2$) of a neutral particle whose radius $\epsilon \to 0$. By the principle of equivalence, this inertial mass $m$ is same as the gravitational mass of the particle. As correctly noted by ADM, in this limit, one has\citep{b1}
\b
m = \lim_{\epsilon \to 0} (32 \pi m_0 \epsilon)^{1/2} =0
\e
independent of the value of the bare mass $m_0$. Essentially, the negative self-gravity completely offsets the original bare mass. The same point has been noted in the classical review paper of ADM\citep{b2} (see p.24):
 
``  Thus as the interaction energy grows more negative, were a point reached where the total energy vanished, there could be no further interaction
energy, in contrast to the negative infinite self-energy of Newtonian theory. General relativity effectively replaces $m_0$ by $m$ in the interaction term: $m=m_0 - (1/2) G m^2/\epsilon$. Solving for $m$ yields $m = G^{-1} [-\epsilon + (\epsilon^2 + 2 G m_0 \epsilon)^{1/2}]$, { which shows that ${\mathbf m \to 0}$ as $\epsilon \to 0$}''.

Therefore, if the proper number density of a supposed `` dust'' be $n(t)$, its proper mass density is still zero:

\b
\rho(t) =\rho_b(t) = m n(t) =0
\e

 In contrast, the proper density of a charged dust however could be finite. This would be so because 

``If a particle is coupled to another field of non-zero range, it may be expected to have non-vanishing total mass due to the interaction with the other field. In particular, by virtue of coulomb field, a point charge $e$ has a total gravitational mass (see Eq.[9]) of \citep{b1}):
\b
m = \gamma^{-1/2} |e|/(4\pi)^{1/2}
\e
where $\gamma =G$. Consequently, a charged ``dust'' may have a finite density unlike a ``neutral dust''.

In the review paper, ADM reconfirm\citep{b2} this result ({\bf see} p.24) and write ``{\em We will see below that this formula is a rigorous consequence of the field equations}''.

\subsection{What about mental picture?}
Many readers may feel uncomfortable with this exact result despite the fact that it has been obtained from both mathematical and physical grounds. This may be so because of the mental picture of spherical dust collapse where point particles are inevitably approaching the geometrical center of symmetry:
\b
{\ddot R} = -{M\over R^2}
\e
  And  such a mental picture about the {\em inevitability} of the dust collapse is partly justified.

When $\rho=M=0$, one has ${\ddot R} =0$ and from Eq.(29), one finds that
 
\b
R(r,t) = f(r) \pm \sqrt{2 E(r)}t
\e
where $2E(r) = -k r^2$. Here we have used the fact that for a uniform density dust, one can write\citep{b28}
\b
R(r,t) = r a(t)
\e
For the collapse case, then one should have
\b
R(r,t) = f(r) - \sqrt{2 E(r)}t
\e
But since $R(r,0) =r$, this becomes
\b
R(r,t) = r - \sqrt{2 E(r)}t = r -r\sqrt{-k}t =r(1-\sqrt{-k}t)
\e
This first rules out any bound $k=+1$ dust. Also for $k=0$, there is no time dependence, no collapse: $a(t) =1$. One can appreciate this by noting that $t_c=\infty$ for $\rho_0 =0$. But for $k=-1$,
 the dust would atleast to form  a special relativistic number density caustic $n=\infty$ at
\b
t =1
\e
Since the spacetime is flat and dust particles have no physical dimension, the converging dust particles must pierce the caustic to expand out.
 For such an assumed expansion, one should have
\b
R(r,t) = f(r) + \sqrt{2 E(r)}t  =f(r) +r t
\e
and this can be recoinciled with $R(r,t) = r a(t)$ form by having
\b
a(t) = t ; \qquad f(r) =0,
\e
and which is alright for an assumed expansion. The line element for this expanding dust becomes\citep{b28}

 \b
ds^2 = dt^2 - t^2 \l[ {dr^2\over 1 +r^2} + r^2 d\Omega^2\r]
\e
 
 As is well known,  by the following coordinate transformations

\b
R = r t
\e
and
\b
T = t\sqrt{1 +r^2},
\e
one can bring the Milne metric into an explicit special relativistic form:
\b
ds^2 = dT^2 - dR^2 - R^2 d\Omega^2
\e

 \section{Summary and Conclusion}
What we have found is that ``dust collapse'' is only an illusion since $\rho_{dust} \equiv 0$.
And accordingly there cannot be any physical singularity associated with a ``dust collapse''.

For a dust ball, this result  could have been anticipated from the following thermodynamic consideration:  For a fluid having finite density, {\em pressure can never be strictly zero}. Conversely, a fluid can strictly have $p=0$ only when $\rho=0$.

   $\bullet$ At a most fundamental level, the result that $\rho(t) =\rho_b= mn =0$ for dust follows from the fact a {\bf neutral point particle has zero gravitational mass}: $m=0$.
Therefore, neither OS nor anybody else have really shown any formation of  black holes out of collapse of a homogeneous dust. This is so even when thousands of authors and scholars have redone the {\em mathematics} of homogeneous dust collapse.  In physics, appearences of mathematical equations need not always reveal the subtle physical reality.  For instance, recently, it was shown that, the maiden interior solution of GR (static uniform density sphere), which has been studied by practically every relativist for the past 94 years is vacuous with $\rho=0$\citep{b23}. 
  
We also found that despite having $\rho=0$, an open homogeneous dust ball can  not only collapse but expand too in the fashion of a Milne universe. But this would be a case of only special relativistic caustic formation.

In retrospect, even if the nuclear fuel would get exhausted for a certain star, {\em internal energy and temperature would not become zero}. Of course, the pressure would reduce and the preexisting  pressure balance would be disturbed. As a result, the star would certainly start contracting. But this does not at all mean that it would get into a free fall. Infact, despite the loss of hydrostatic balance, neither pressure, nor temperature nor internal energy would ever strictly become zero. While the star would certainly undergo gravitational contraction if the nuclear fuel of the star would be exhausted, in view of the   {\em negative specific heat} associated with a self- gravitating fluid, the star would actually become hotter! And this would create a tendency for generating extra pressure\citep{b16}. Indeed, in reality, all gravitational collapse processes are bound to be dissipative and radiative\citep{b17}. One of the best examples of it is SN1987A which radiated $\sim 4\times 10^{53}$ erg in neutrinos alone. In fact, it has been argued that, if the collapsing object would dive deeper into the gravitational potential well, the contracting and hot object would
release even more energy.  A generic study has shown that  a contracting body on its way to continued collapse must tend to be radiation pressure dominated\citep{b18}. Several studies have also suggested that the contracting body may even radiate out its entire mass energy to avoid formation of trapped surfaces or horizons\citep{b3,b5,b27}.

 It has also been pointed out that, as the collapsing object would dive into its photon sphere at $z > \sqrt{3} -1$ it would start trapping its own radiation. And in view of increase of temperature as well as trapping of radiation by its own gravity, general relativistic continued collapse may result in a new quasi-static state when the surface gravitational redshift of the object $z\gg 1$\citep{b19,b22}. Such radiation pressure supported quasi-static objects are {\em just the extreme general relativistic versions ($z\gg 1$) of quasi-Newtonian} ($z \ll 1$) radiation pressure supported objects {\bf conceived by Hoyle and Fowler} long back\citep{b7,b9}. Indeed, it is known that GR may allow extremely compact radiation dominates massive spheres too and which can be viewed as the central engine of Active Galactic Nuclei\citep{b25,b26,b31}. In particular, there are specific model for formation of such radiation supported objects in {\bf horizonless} GR collapse\citep{b26}.

Quite ironically, though the OS paper is considered as the bible for theoretical proof for formation of black holes,   everybody seems to have forgot what OS wrote before they proceeded to offer their solution:

$\bullet$

``Physically such a singularity would mean that the expressions used for the energy-momentum tensor does not take into account  some {\em essential physical fact} which would really {\em smooth the singularity out}. Further, a star in its early stages of development would not possess a singular density or pressure, it is impossible for a singularity to develop in a finite time.'' (see p.456, column 2\citep{b24}).

And here we showed that, for the OS collapse, indeed  {\em the energy momentum tensor does not take into account  some essential fact}; and the fact is that for a dust particle $m=0$.
And when one would take this into account, this fact will indeed ``smooth the singularity out'' -- the same singularity suggested by mathematics and brusing aside nuances of physics!

And the conclusion reached in this paper that, in reality, OS work did not show the formation of any {\em finite mass exact black hole} is in agreement with many previous theoretical results that GR prohibits formation of {\bf exact} black holes\citep{b15,b4, b10, b12, b29, b30}. It might be meaningful to recall the title of these last four references  

1. ``Black holes: a prediction of theory or fantasy?''

2. ``The physical inconsistency of the Schwarzschild and Kerr solutions''

3. `` General relativity and conformal invariance. I - A new look at some old field equations.''

and 

4. ``General relativity and conformal invariance. II. Non-existence of black holes''.


\end{document}